\documentclass[doublecol]{epl2} 

\usepackage{amsmath}

\newcommand{\bg}{\bar{\gamma}}

\newcommand{\enc}{\mathrm{enc}}

\newcommand{\rmi}{\mathrm{i}}

\newcommand{\rme}{\mathrm{e}}
\newcommand{\diag}{\mathrm{diag}}

\newcommand{\UI}{\mathrm{I}}

\title{Periodic orbit evaluation of a spectral statistic of quantum graphs without the semiclassical limit}
\shorttitle{Periodic orbit evaluation without the semiclassical limit} 

\author{J.M. Harrison\inst{1} \and T. Hudgins\inst{2}}
\shortauthor{J.M. Harrison and T. Hudgins}

\institute{                    
  \inst{1} Baylor University - Department of Mathematics,
  1410 S. 4th Street, Waco, TX 76706, USA\\
  \inst{2} University of Dallas - Department of Mathematics,
  1845 E. Northgate Dr., Irving, TX 75062, USA
}

\abstract{
	Energy level statistics of quantized chaotic systems have been evaluated in the semiclassical limit via their periodic orbits using the Gutzwiller and related trace formulae.
	Here we evaluate a spectral statistic of chaotic $4$-regular quantum graphs from their periodic orbits without the semiclassical limit.   The variance of the $n$-th coefficient of the characteristic polynomial is determined by the sizes of the sets of distinct primitive periodic orbits with $n$ bonds which have no self-intersections, and the sizes of the sets with a given number of self-intersections which all consist of two sections of the pseudo orbit crossing at a single vertex.
	Using this result we observe the mechanism that connects semiclassical results to the total number of orbits regardless of their structure.}

\begin{document}

\maketitle

\section{Introduction}

The analysis of chaotic quantum systems via their periodic orbit structure originates with Gutzwiller's trace formula \cite{Gutzwiller}.  It has been widely applied to analyze spectral statistics like the form factor (the Fourier transform of the two-point correlation function) in the semiclassical limit.   This starts with Berry \cite{B85} who evaluates a diagonal contribution, pairing a periodic orbit with an identical partner orbit, using the sum rule of Hannay and Ozorio de Almeida \cite{HA84}.  More recently Sieber and Richter \cite{S02,SR01} evaluated a first order correction in the semiclassical limit by considering contributions from figure eight orbits with a self-intersection where a second partner orbit of the same length can be generated by reordering the order that loops in the figure eight are traversed.  This scheme was extended to evaluate all higher order corrections by M\"uller et. al. \cite{Metal04,Metal05}.  These are only a few of the results obtained in this manner.

Quantum graphs are used to study quantum systems with complex topology  in fields from quantum chemistry and nanotechnology, to waveguides, Anderson localization and carbon nanostructures, see \cite{BerkolaikoKuchment} for an introduction.  In \cite{KS97,KS99} Kottos and Smilansky introduced quantum graphs to study quantum mechanics in a system that is classically chaotic, see \cite{GS06} for a review.  In particular, they demonstrated that, in the semi-classical limit, spectral statistics can be traced to topological properties of the periodic orbits.   Quantum graphs provided an important model where arguments to extend the semiclassical evaluation of the form factor were developed \cite{BSW02,BSW03}.
One dimensional network models were recently used to investigate quantum signatures of chaos outside the regime of universality in \cite{Fetal20}.
Spectral properties of quantum graphs have also been investigated experimentally using microwave networks \cite{Hetal04,JMS14,Retal16}.  

We show that, for a large class of chaotic $4$-regular quantum graphs, one can precisely evaluate a periodic orbit formula for a spectral statistic, the variance of coefficients of the characteristic polynomial of the quantum evolution operator, from the periodic orbit structure without the semiclassical limit.  This variance was first considered in the initial results of Kottos and Smilansky \cite{KS99}.  Semiclassical results for the variance were produced by Tanner  for binary quantum graphs, which are a class of $4$-regular graphs, using a diagonal approximation \cite{T02}.  The same system was also analyzed semiclassically in \cite{BHS19}  by evaluating a diagonal contribution to a pseudo orbit formula. 
The result we present is surprising as we see that the variance is determined only by periodic orbits with certain structures while the semiclassical results weight all orbits equally.  Taking a semiclassical limit in the precise variance formula we observe the mechanism that allows a diagonal approach to produce the correct semiclassical result.

\section{Graph model}

\begin{figure}[htbp!]
	\includegraphics[scale=0.79, trim=145 570 150 125, clip]{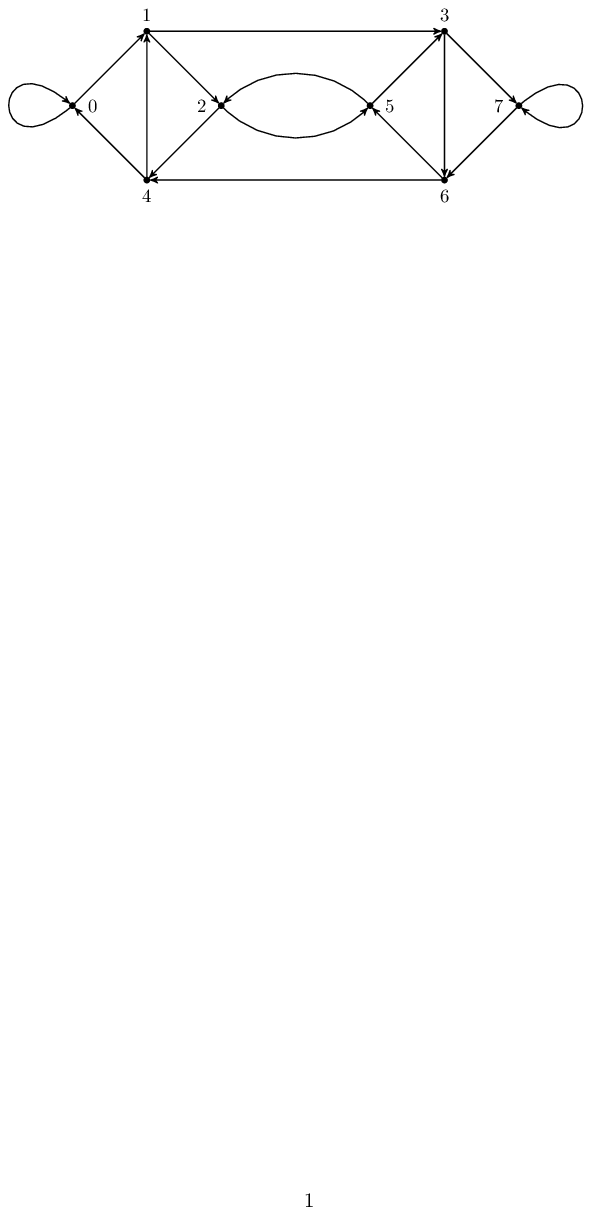}
	\caption{Binary de Bruijn graph with $2^3$ vertices.}
	\label{fig: binary V=8}
\end{figure}

A graph is a network of $V$ vertices connected by $B$ bonds.  We consider $4$-regular graphs, graphs where every vertex is connected by a bond to exactly four neighboring vertices.  
We take the graph to be directed, so every vertex has two bonds that are incoming and two that are outgoing.  
This is always possible as a $4$-regular connected graph admits an Euler tour.
Fig. \ref{fig: binary V=8} shows such a $4$-regular directed graph with $V=8$ vertices.  A bond $b=(u,v)$ is an ordered pair of vertices; $o(b)=u$ and $t(b)=v$, the origin and terminus of $b$, respectively.
In a quantum graph the bonds are assigned a positive length which we collect in a diagonal matrix $L=\diag\{ L_1,\dots,L_{B} \}$.   At each vertex $v$ there is an associated unitary vertex scattering matrix $\sigma^{(v)}$ which describes scattering from incoming to outgoing bonds at $v$. The elements of these matrices are collected in a $B\times B$ unitary bond scattering matrix, 
\begin{equation}
\label{eq:bondSmatrix}
S_{b',b} = \delta_{o(b'),v}\ \delta_{t(b),v}\   \sigma_{b',b}^{(v)} \ . 
\end{equation}  
The spectrum of a quantum graph corresponds to roots of a secular equation,
\begin{equation}
\label{eq:secular}
\det(S \rme^{\rmi k L}-\UI)=0 \ ,
\end{equation}
where $k$ is the spectral parameter \cite{KS97,KS99}.  
The matrix $U(k)=S \rme^{\rmi k L}$ is the quantum evolution operator for the graph. 

Given a Hamiltonian on a metric graph, the vertex scattering matrices are determined by the matching conditions at the vertices.  If $k\neq 0$ is a root of the secular equation (\ref{eq:secular}) then $k^2$ is an eigenvalue of the operator with the same multiplicity.  Here we adopt a complementary approach to quantize graphs where we simply specify unitary scattering matrices at the vertices \cite{T01}.  This defines a unitary stochastic matrix ensemble $U(k)$.  The vertex scattering matrix that we use to quantize $4$-regular graphs is the discrete Fourier transform (DFT) matrix,
\begin{equation}
\sigma^{(v)} = \frac{1}{\sqrt{2}}  
\begin{pmatrix}
1 & 1 \\
1 & -1 \\
\end{pmatrix}
\ .
\label{eq:DFTmatrix}
\end{equation}
The DFT matrix is frequently used to quantize graphs \cite{GS06}.  It has the advantage that it is democratic, in the sense that the transmission probability $|\sigma_{b',b}^{(v)}|^2$ is the same for every pair $b,b'$. 

\section{Characteristic polynomial}

The characteristic polynomial of a quantum graph is,
\begin{equation} 
\label{eq:charpoly}
\det(U (k) - \zeta \UI) = \sum\limits_{n=0}^{B} a_n \zeta^{B-n} \ .
\end{equation}
Consequently, the left hand side of the secular equation (\ref{eq:secular}) is the characteristic polynomial for $\zeta=1$.   The unitarity of $U(k)$ produces a Riemann-Siegel lookalike formula for the coefficients $a_n=a_B\bar{a}_{B-n}$ \cite{KS99}.  

The $n$-th coefficient can be written as a sum over sets of distinct primitive periodic orbits (primitive pseudo orbits) \cite{BHJ12}.  A periodic orbit is an equivalence class of closed cycles, where cycles are equivalent if they are related by a cyclic permutation.  A periodic orbit is primitive if it is not a repetition of a shorter orbit.  
Averaging over $k$, the spectral parameter, $\langle a_n \rangle=0$ for $0<n<B$ ($\langle a_0 \rangle=\langle a_B \rangle=1$) and the first non-trivial moment is the variance,
\begin{equation}
\langle |a_n|^2 \rangle
= \sum_{\bg, \bg' \in \mathcal{P}^n} 
(-1)^{m_{\bg}+m_{\bg'}} 
A_{\bg} \bar{A}_{\bg'}   
\delta_{L_{\bg}, L_{\bg'}} \ ,
\label{eq:variance sum} 
\end{equation}		
where $\mathcal{P}^n$ is the set of primitive pseudo orbits with $n$ bonds.  A primitive pseudo orbit $\bg=\{\gamma_1, \gamma_2, \dots, \gamma_{m_{\bg}}\}$ 
consists of $m_{\bg}$ distinct primitive periodic orbits $\gamma_i$, so $\gamma_i \neq \gamma_j$ for $i\neq j$.    Another class of pseudo orbits that will be relevant are irreducible pseudo orbits.  These are pseudo orbits where all of the bonds in the pseudo orbit are distinct.  Irreducible pseudo orbits are a subset of primitive pseudo orbits as if an orbit is repeated in a pseudo orbit or a pseudo orbit contains a nonprimitive periodic orbit the pseudo orbit is not irreducible.   

Associated to a periodic orbit $\gamma$ is a stability amplitude $A_{\gamma}$, the product of the scattering amplitudes of the vertex scattering matrices associated with $\gamma$.  For a pseudo orbit, the stability amplitude $A_{\bg}$ is the product of the stability amplitudes of the orbits in $\bg$. 
The length of a periodic orbit $L_{\gamma}$ is the sum of the lengths of the bonds in $\gamma$, and $L_{\bg}$ is the sum of the lengths of the orbits in $\bg$.  Notice that pairs of primitive pseudo orbits $\bg, \bg'$ that contribute to the variance (\ref{eq:variance sum}) must have the same length $L_{\bg}=L_{\bg'}$.
We take the set of bond lengths to be incommensurate, so if two pseudo orbits have the same length they must visit each bond the same number of times.  
Given a pseudo orbit $\bg$, to generate a different pseudo orbit $\bg'$ that visits the same set of bonds, $\bg$ must have self-intersections.  

\begin{figure}[htbp!]
	\includegraphics[scale=0.77, trim=115 520 175 125, clip]{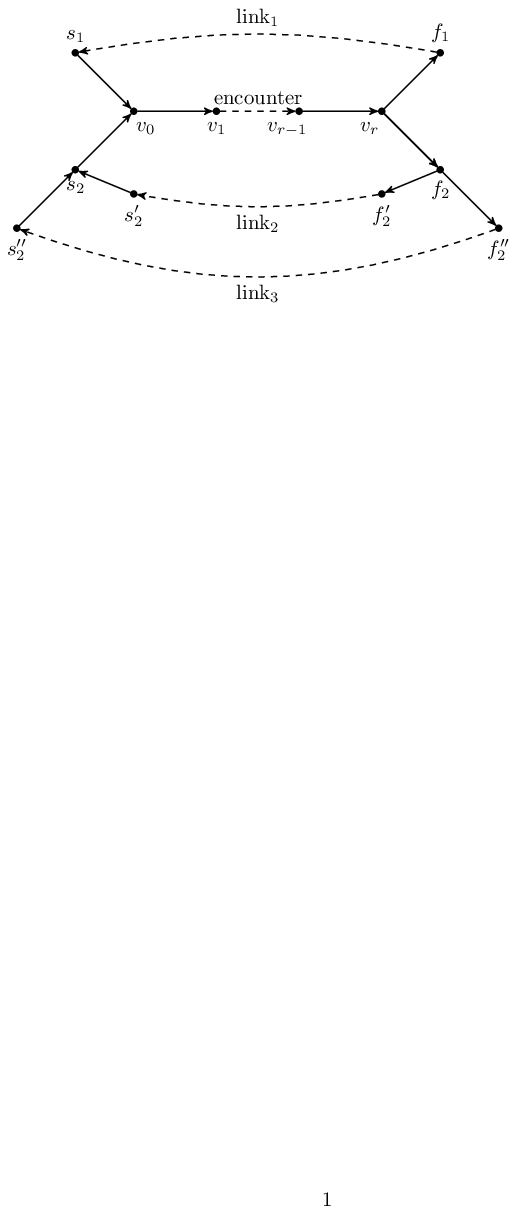}
	\caption{Subgraph on which pseudo orbits with self-intersections can be constructed.} 
	\label{fig:encounter}
\end{figure}

Fig. \ref{fig:encounter} shows a subgraph on which one can construct pseudo orbits with self-intersections.  In the figure, the sequence of adjacent vertices $\enc=(v_0,\dots,v_{r})$ will be the encounter which is the maximally repeated section of the pseudo orbit inside the self-intersection.   The encounter length will be the number of bonds in the encounter $r$.  An encounter can take place at a single vertex $\enc=(v_0)$ in which case the encounter has length zero.  When the encounter sequence appears $l$ times in the pseudo orbit we say that it is an $l$-encounter.

Sequences of adjacent vertices joining the end of an encounter to the start of an encounter are called links.  Pseudo orbits can be defined by specifying the sequences of links in each orbit.  For example, consider a primitive pseudo orbit $\bg=(\gamma_1,\dots, \gamma_m)$ where there are no self-intersections in $\gamma_2,\dots,\gamma_m$ and $\gamma_1 = (1 \, 2)$, by which we mean the orbit follows link $1$, followed by the encounter, then link $2$, followed by the encounter a second time, i.e.,
\begin{equation}
\gamma_1 = (f_1 \dots, s_1, \enc, f_2, f_2' \dots, s_2', s_2, \enc , f_1) \ ,
\label{eq:2enc}
\end{equation}
so $\bg$ has a $2$-encounter.  Similarly, we could define a pseudo orbit with a $3$-encounter by taking $\gamma_1=(1\,2\,3)$.  In this case there are bonds preceding and following the encounter that are repeated twice, $(s_2,v_0)$ and $(v_r,f_2)$, which is necessary to produce a $3$-encounter on a directed $4$-regular graph.  In either of the previous examples, if $v_0=v_r$, there are no bonds in the encounter and we have an encounter of length zero.  However, while the pseudo orbit $(1\,2)$ has no repeated bonds and so is irreducible the pseudo orbit $(1\,2\,3)$ still requires the two bonds $(s_2,v_0)$ and $(v_r,f_2)$ preceding and following the encounter to be repeated even when the encounter has length zero, and so it is not irreducible.  

The previous examples have an encounter repeated in a single periodic orbit.  However, we could instead define a pseudo orbit with a $2$-encounter by taking $\gamma_1=(1)$ and $\gamma_2=(2)$ with the remaining orbits not intersecting any other orbit.  In this case both orbits follow a link once and then the encounter once.
Finally, we also point out that pseudo orbits with an $l$-encounter with $l>2$ can repeat links. So, for example, $\gamma_1=(1\,2\,3\,2)$ defines an orbit with a $4$-encounter which uses link $2$ twice.

Given a pseudo orbit with one or more encounters one can obtain partner pseudo orbits of the same length by rearranging the order of the links at the encounters.  For example given a primitive pseudo orbit $\bg$ with $\gamma_1 = (1 \, 2)$ and no other self-intersections there are two primitive partner pseudo orbits of the same length $\bg'=\bg$ and $\bg'= ((1),(2),\gamma_2,\dots, \gamma_m)$.  Similarly if the primitive pseudo orbit $\bg$ includes a $3$-encounter, for example $\gamma_1=(1\,2\,3)$, then there are six partner pseudo orbits we can obtain by replacing $\gamma_1$ with an element of the set of pseudo orbits,
\begin{equation*}
\{ (1)(2)(3), (12)(3), (13)(2), (1)(23), (123), (132) \} \ .
\end{equation*} 
One of these partners is again the pseudo orbit $\bg$ itself and evaluating the contribution to equation (\ref{eq:variance sum}) obtained by setting $\bg'=\bg$ was the method used to obtain the semi-classical result for families of graphs with increasing numbers of bonds \cite{BHJ12,T02}.

\section{Evaluation of the variance}

We now state the main observation,
\begin{equation}
\label{eq:variancefinal}
\langle |a_n|^2 \rangle
= \frac{1}{2^{n}} \left( |\mathcal{P}_0^n| + \sum_{j=1}^n 2^j \, |\widehat{\mathcal{P}}_{j}^n| \right) \ ,
\end{equation}
where $\mathcal{P}_0^n$ is the set of primitive pseudo orbits of length $n$ with no self-intersections and $\widehat{\mathcal{P}}_{j}^n$ is the set of primitive pseudo orbits of length $n$ with $j$ self-intersections, all of which are $2$-encounters of length zero.  The sum is finite and we can use $n$ as the last term in the sum, as an orbit of length $n$ clearly cannot have more than $n$ self-intersections.

To demonstrate the result, tables \ref{Table binary 8} and \ref{table binary 6} compare the variance of the first half of the coefficients of the characteristic polynomial evaluated using (\ref{eq:variancefinal}) with the numerically computed variance for two binary quantum graphs.  The variance of the remaining coefficients is determined by the symmetry about $n=B/2$.
Binary graphs are $4$-regular graphs { with $V=p2^s$ vertices and $B=p2^{s+1}$ bonds} where $p$ is odd.  They were studied by Tanner in this context \cite{T00,T01,T02} where only asymptotic results are obtained.   The bonds of binary graphs are specified by the adjacency matrix,
\begin{equation} 
\label{qnaryadjacency}
[A]_{ij} = 
\begin{cases} 
\delta_{2i,j}+\delta_{2i+1,j}, & 0\leq i <V/2 \ ,\\
\delta_{2i-V,j}+\delta_{2i+1-V,j} & V/2 \leq i < V  \ ,
\end{cases}
\end{equation}
so $[A]_{ij}=1$ when $i$ is the origin of a bond whose terminus is $j$ and zero otherwise. The binary de Bruijn graphs are the family of binary graphs with $p=1$. Fig. \ref{fig: binary V=8} shows the binary de Bruijn graph with $V=2^3$ and table \ref{Table binary 8} shows the values of the variance of the coefficients of the characteristic polynomial of this graph. 
The numerical averages over $k$ were computed for intervals of the spectrum containing 50 million eigenvalues.  
The sets of primitive pseudo orbits used to compute the variance from (\ref{eq:variancefinal}) are in \cite{HH20}.  In each table we see agreement to at least four decimal places.

\begin{table}[htbp!]
	\caption{Variance of coefficients of the characteristic polynomial of a binary de Bruijn quantum graph with $2^3$ vertices.}
	\label{Table binary 8}
	\begin{center}{\small
	\begin{tabular}{ c c c c c c c }
		$n$ & $|\mathcal{P}_0^n|$ & $|\widehat{\mathcal{P}}_{1}^n|$ 
		& $|\widehat{\mathcal{P}}_{2}^n|$ & $\langle |a_n|^2 \rangle$ 
		& Numerics & Error \\
		\hline
		0 & 1 & 0 & 0 & 1 & 1.000000 & 0.000000 \\
		1 & 2 & 0 & 0 & 1 & 0.999991 & 0.000009 \\
		2 & 2 & 0 & 0 & 1/2 & 0.499999 & 0.000001 \\
		3 & 4 & 0 & 0 & 1/2 & 0.499999 & 0.000001 \\
		4 & 8 & 0 & 0 & 1/2 & 0.499999 & 0.000001 \\
		5 & 8 & 8 & 0 & 3/4 & 0.749998 & 0.000002 \\
		6 & 8 & 20 & 0 & 3/4 & 0.749986 & 0.000014 \\
		7 & 16 & 16 & 8 & 5/8 & 0.624989 & 0.000011 \\
		8 & 16 & 16 & 24 & 9/16 & 0.562501 & -0.000001 \\
		\hline
	\end{tabular}}
\end{center}
\end{table}

\begin{table}[htbp!]
	\caption{Variance of coefficients of the characteristic polynomial of a binary quantum graph with $3\cdot 2$ vertices.}
		\label{table binary 6}
		\begin{center}
	\begin{tabular}{ c c c c c c }
		$n$ & $|\mathcal{P}_0^n|$ & $|\widehat{\mathcal{P}}_{1}^n|$ 
		& $\langle |a_n|^2 \rangle$ 
		& Numerics & Error \\
		\hline
		0 & 1 & 0 & 1 & 1.000000 & 0.000000 \\
		1 & 2 & 0 & 1 & 1.000000 & 0.000000 \\
		2 & 3 & 0 & 3/4 & 0.750001 & -0.000001 \\
		3 & 6 & 0 & 3/4 & 0.750003 & -0.000003 \\
		4 & 10 & 4 & 7/8 & 0.874999 & 0.000001 \\
		5 & 8 & 4 & 1/2 & 0.499998 & 0.000002 \\
		6 & 8 & 8 & 3/8 & 0.374999 & 0.000001 \\
		\hline
	\end{tabular}
\end{center}
\end{table}

\section{Semiclassical limit}

{ Spectral statistics of quantum graphs can be formulated dynamically in terms of their periodic orbits or pseudo orbits, as in (\ref{eq:variance sum}), as the trace formula for quantum graphs is exact.  However, to evaluate these expressions typically requires using approximations valid in the semiclassical limit.}
The set of $4$-regular quantum graphs provides a large class of quantum chaotic systems with a quantum statistic that can be computed dynamically and precisely without the semiclassical limit.
The form of the result is somewhat surprising as only particular sets of pseudo orbits contribute, while, in the semiclassical limit, the asymptotic variance can be obtained using diagonal arguments which weight all primitive pseudo orbits \cite{BHS19} or all orbits \cite{T02} equally.  The semiclassical limit is the limit of increasing spectral density, which for quantum graphs is the limit of a sequence of graphs with an increasing number of bonds \cite{BerkolaikoKuchment,KS97,GS06}.  For families of binary graphs with $p$ fixed and $B=p2^{s+1}$ the semiclassical limit is $s\to \infty$.   To take the semiclassical limit in  (\ref{eq:variancefinal}) we can determine the proportion of orbits where all the self-intersections are  $2$-encounters of length zero on large graphs, fixing the ratio $n/B$.    

On $4$-regular directed graphs if we follow a pseudo orbit with a $2$-encounter, when we reach the initial encounter vertex for the second time, there will be two outgoing bonds, one of which has already appeared in the orbit.  The probability to scatter in each direction is $1/2$ and so, for long pseudo orbits on a large graph, half of primitive pseudo orbits with a $2$-encounter will use the bond that has not appeared in the orbit already; a $2$-encounter of length zero.  If, instead of a $2$-encounter, the pseudo orbit has a $3$-encounter the pseudo orbit must use the encounter vertex twice and then hit the same vertex a third time.  As the graph is mixing \cite{KS99}, the probability to end on any bond after a large number of steps is $1/B$.  Hence, for long orbits on large graphs, the number of $3$-encounters in pseudo orbits is vanishingly small compared to the number of $2$-encounters.  Let $\mathcal{P}^n_j$ denote the set of primitive pseudo orbits of length $n$ with $j$ encounters.  Then
\begin{equation}\label{eq: asymp ratio}
|\widehat{\mathcal{P}}_{j}^n|\approx 2^{-j} |\mathcal{P}^n_j| \ .
\end{equation}
Consequently from (\ref{eq:variancefinal}),
\begin{equation}
\label{eq:varianceasmp}
\langle |a_n|^2 \rangle \approx 2^{-n} \sum_{j=0}^n |\mathcal{P}^n_j| = 2^{-n}\, |\mathcal{P}^n|  \ .
\end{equation}
So asymptotically the variance is proportional to the total number of primitive pseudo orbits of length $n$. 

\begin{figure}[htbp!]
	\includegraphics[scale=0.69, trim=150 492 100 125, clip]{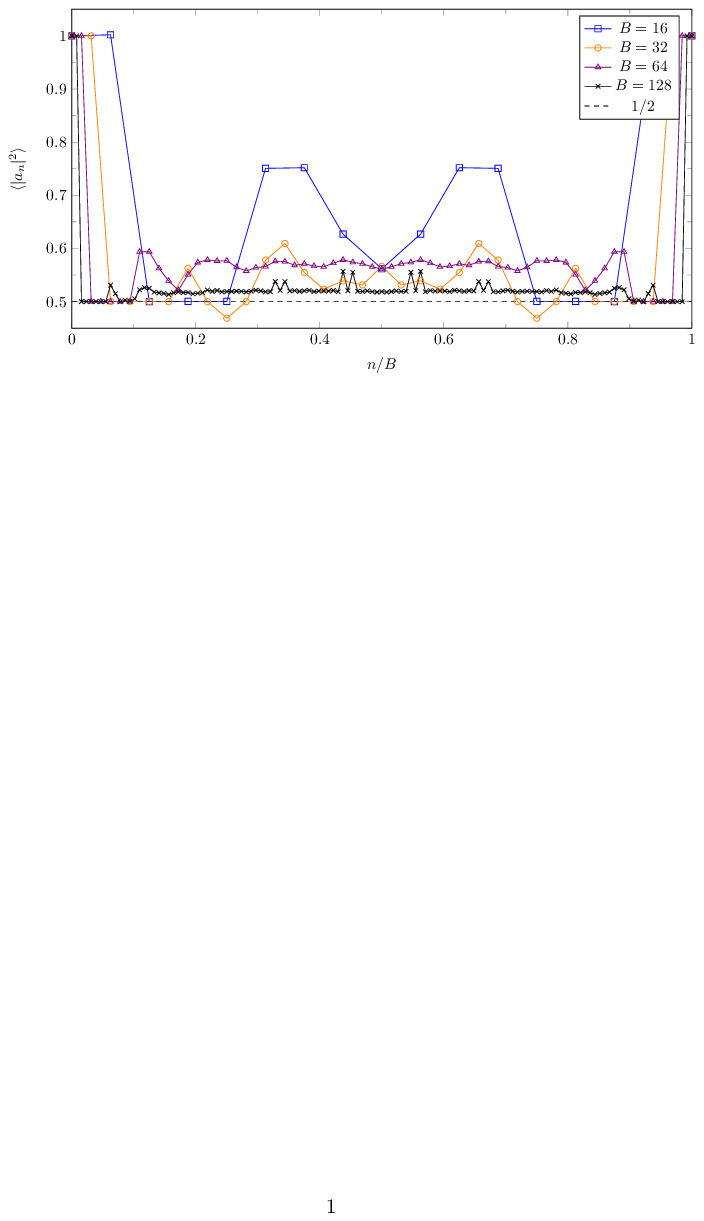}
	\caption{Variance of coefficients of the characteristic polynomial of binary de Bruijn quantum graphs showing convergence to the asymptotic value $1/2$.} 
	\label{fig: p=1 family}
\end{figure}

For binary de Bruijn graphs the number of primitive pseudo orbits of length $n$ is $2^{n-1}$ \cite{BHS19} and hence $\langle |a_n|^2 \rangle\approx 1/2$, which agrees with the result obtained by evaluating a diagonal contribution \cite{T02, BHS19}.
Fig. \ref{fig: p=1 family} shows the convergence of the variance to the asymptotic result for the family of binary de Bruijn graphs.

\section{Derivation of the variance formula}

Equation (\ref{eq:variancefinal}) can be obtained in multiple ways.  Here we describe a short argument exploiting cancellations in the pseudo orbit sum identified in \cite{BHJ12}.   The formula (\ref{eq:variancefinal}) can also be produced directly using a parity argument based on Lyndon tuples which offers the possibility that results may be able to be extended beyond these quantum graph models, see \cite{HH20} .  

In \cite{BHJ12} Band, { Harrison} and Joyner showed that the sum over the set of primitive pseudo orbits (\ref{eq:variance sum}) can be reduced to a sum over the set of irreducible pseudo orbits; those pseudo orbits where no bond in the graph appears more than once, 
\begin{equation}
\langle |a_n|^2 \rangle
= \sum_{\bg, \bg' \in \widehat{\mathcal{P}}^n} 
(-1)^{m_{\bg}+m_{\bg'}} 
A_{\bg} \bar{A}_{\bg'}   
\delta_{L_{\bg}, L_{\bg'}} \ ,
\label{eq:irreducible variance sum} 
\end{equation}
where $\widehat{\mathcal{P}}^n$ is the set of irreducible pseudo orbits of length $n$.
Starting from this formula, we immediately see that if an irreducible pseudo orbit has encounters then all the encounters must be of length zero, otherwise there are repeated bonds in the pseudo orbit.  In addition, in order to have a $3$-encounter of length zero at least two bonds must be repeated in the pseudo orbit when entering and exiting the encounter vertex.  This also rules out $l$-encounters for $l\geq 3$.  Hence the only type of encounter that can be used to reorder links and produce partner irreducible pseudo orbits $\bg'$ of the same length as $\bg$  in (\ref{eq:irreducible variance sum}) are $2$-encounters of length zero.  It remains to evaluate the contribution generated by such partner pseudo orbits.

To evaluate the double sum in (\ref{eq:irreducible variance sum}) we define a contribution for each irreducible pseudo orbit $\bg$,
\begin{equation}
C_{\bg} = \sum_{\bg' \in \widehat{\mathcal{P}}_{\bg} }
(-1)^{m_{\bg}+m_{\bg'}}
A_{\bg} \bar{A}_{\bg'} \ ,
\label{eq:C}
\end{equation}
where $\widehat{\mathcal{P}}_{\bg}$ is the set of irreducible pseudo orbits of length $L_{\bg}$. 
The variance of the coefficients is then,
\begin{equation}
\label{eq:variance C sum}
\langle |a_n|^2 \rangle =
\sum_{\bg \in \widehat{\mathcal{P}}^n} C_{\bg} \ .
\end{equation}

The first term in (\ref{eq:variancefinal}) is now straightforward to explain.  From (\ref{eq:DFTmatrix}) we see that, for a pseudo orbit of $n$ bonds, $|A_{\bg}|^2=1/2^n$.  If $\bg$ is an irreducible pseudo orbit with no self-intersections it is not possible to reorder the bonds of the pseudo orbit to produce a different orbit with the same length.  Hence ${ \widehat{\mathcal{P}}_{\bg}}=\{\bg \}$ and $C_{\bg}=1/2^n$.  
Finally, any primitive pseudo orbit that is not irreducible has a self-intersection as it has a repeated bond, so the set of primitive pseudo orbits with no self-intersections is the set of irreducible pseudo orbits with no self-intersections.

The remaining irreducible pseudo orbits are those with self-intersections where all the self-intersections are $2$-encounters of length zero. Let $j$ be the number of $2$-encounters and $i$ the number of those encounters at which links in the partner orbit $\bg'$ are reordered relative to $\bg$.  Each encounter at which $\bg'$ is reordered 
changes the number of periodic orbits in the pseudo orbit by $\pm 1$.  Hence $(-1)^{m_{\bg'}}=(-1)^{m_{\bg}+i}$.  Also at every encounter where $\bg'$ is reordered there is an additional change in the sign of the stability amplitude, so $A_{\bg} \bar{A}_{\bg'} =(-1)^i 2^{-n}$.  Consequently,
\begin{equation}
C_{\bg} =  
2^{-n}
\sum_{i=0}^j (-1)^{2i}
\binom{j}{i}   
= 2^{j-n}  \ .
\end{equation}
Substituting these in (\ref{eq:variance C sum}) produces (\ref{eq:variancefinal}) as all primitive pseudo orbits that are not irreducible have a repeated bond and so have at least one encounter of length greater than zero.

\section{Conclusion} In summary, we have demonstrated that for a large class of chaotic $4$-regular quantum graphs there is a non-trivial spectral statistic that can be evaluated dynamically without taking the semiclassical limit.   The variance of coefficients of the characteristic polynomial is precisely determined by the sizes of sets of distinct primitive periodic orbits with  particular structures; those sets of orbits where all the self-intersections are $2$-encounters of length zero.  This is an exciting development as it provides the first example of a chaotic quantum system where the dynamical formulation of a spectral statistic can be evaluated without taking the semiclassical limit.   While the importance of orbits with self-intersections is well established in the semiclassical arguments their contribution here is novel, as only orbits whose self-intersections have a particular structure contribute. Alternative parity arguments based on Lyndon tuples \cite{HH20} that produce the same result suggest that it may be possible to  extend the result beyond this quantum graph model.  

For $4$-regular graphs we can also see the mechanism that allows semiclassical arguments based on a diagonal approach to produce the correct result.   
 It turns out that the weight $2^j$, on the size of the set of primitive pseudo orbits with $j$ self-intersections which are all $2$-encounters of length zero in (\ref{eq:variancefinal}), is asymptotically the ratio of the size of that set to the size of the set of all primitive pseudo orbits with $j$ self-intersections of any type,
\begin{equation}\label{eq: ratio}
2^j\approx  \frac{|\mathcal{P}^n_j|}{|\widehat{\mathcal{P}}_{j}^n|} \  ,
\end{equation}
see (\ref{eq: asymp ratio}).

\acknowledgments
The authors would like to thank Gregory Berkolaiko for helpful comments.
JH would also like to thank Mark Pollicott and the University of Warwick for their
hospitality during his sabbatical where JH was
supported by the Baylor University research leave program. This work was 
supported by a grant from the Simons Foundation (354583 to Jonathan Harrison).

\end{document}